# Unique catalytic decomposition and reduction of hydrogen peroxide and organic peroxides on magnesia (001) promoted by oxide-metal hybrid nanostructure

Zhenjun Song[a], Bin Zhao[a,*], Qiang Wang[b], and Peng Cheng[a,*]

[a] *Department of Chemistry, Key Laboratory of Advanced Energy Material Chemistry (Ministry of Education) and Collaborative Innovation Center of Chemical Science and Engineering (Tianjin), Nankai University, Tianjin 300071, People's Republic of China.*

[b] *State Key Laboratory of Coal Conversion, Institute of Coal Chemistry, Chinese Academy of Sciences, Taiyuan 030001, Shanxi, People's Republic of China.*

**ABSTRACT:** The detection, removal and reduction of hydrogen peroxide is of significant importance for its increasing application in the areas of environment, food, electrochemistry and clinical laboratory. Herein the dissociative adsorption behavior of hydrogen peroxide on ultrathin magnesia (001) films deposited on transition metal is uncovered for the first time by employing periodic density-functional theory calculations with van der Waals corrections. The dissociation of hydrogen peroxide on bulk MgO(001) is calculated to be highly endothermic process with activation barrier 1.85 eV, indicating it is extraordinarily difficult to dissociate hydrogen peroxide on pristine MgO(001). The hydrogen peroxide is dissociated smoothly and reduced to surface hydroxyls on MgO(001)/TM, and the dissociative adsorption energies of all the considered fragmentation configurations are substantially negative, demonstrating dissociation and reduction of hydrogen peroxide on MgO(001)/TM is thermodynamically favorable. From the comparison between the dissociation behavior on bare magnesia, extended bare magnesia, and metal-supported magnesia, it can be deduced that the metal substrate should play a crucial role in dissociation and reduction of hydrogen peroxide. The dissociative adsorption energy decreases monotonously with increasing film thickness, demonstrating the lower reactivity of thick oxide films. Moreover, we examined the suitability of several transition metal slabs (molybdenum, silver, vanadium, tungsten and gold) combined with magnesia for splitting hydrogen peroxide. The dissociative adsorption energies enhance when the lattice constants of substrate slabs increase, indicating the chemisorption strength can be tuned by category of metal slabs as well as thickness of oxide film. The mechanism of reactivity enhancement for energetically and dynamically favorable decomposition of hydrogen peroxide on supported magnesia is elucidated by characterizing the geometric structures and electronic properties. The interaction between hydrogen peroxide and ultrathin magnesia covered with water/methanol is considered to investigate the influence of solvent molecules. The fragmentation and reduction of diethyl peroxide and peroxyacetone are also studied to reveal the catalytic activity of ultrathin magnesia toward splitting organic peroxides. The results are wished to provide useful clue for detecting and reducing hydrogen peroxide and organic peroxides by employing oxide-metal hybrid nanostructure.

## 1. Introduction

Metal oxides and oxide-based composite nanostructures have been researched extensively for potential technical applications and reported to be efficient nanocatalyst for many significant tasks such as transformation of carbon dioxide into hydrocarbon compounds, low-temperature oxidation of carbon monoxide, photocatalysis, solar energy conversion et al.[1-7] In contrast to the high reactivity of reducible metal oxides,[8-10] the perfect and nonpolar MgO(001) surface with chemical inertness are generally believed to be inactive and rarely investigated for catalyzing challenging reactions. The low coordinated sites such as defects, steps and corners are usually confirmed to control the catalytic reaction on MgO(001) surface.[11] With the development of advanced lithographic and imprinting techniques, the self-assembly and patterning of oxides can be accomplished at nanometer.[12] Pacchioni and Giordano reviewed spectacular advances of functional oxide films at the nanoscale and proposed that the ultrathin oxide films could provide tremendous potential and unforeseen opportunities for heterogeneous catalysis.[13] The active sites of heterogeneous catalyst are distributed on ultrathin oxide films, rather than the bare metal. Although the insulating magnesia support is usually ascertained to be chemically inert, when the dimensionality of magnesia (such as thickness of oxide film) enters nanometer regime, they exhibit extraordinary properties without resembling the bulk oxide. Experimentally, the ultrathin magnesia film can be obtained by reactive deposition of Mg on sputtered and annealed molybdenum or silver surface in an oxygen atmosphere.[14] The coincidence lattice and electronic modulation between oxide and metal provides valuable approach to realize preferred nucleation and growth of admetals. In scanning tunneling microscopy experiments, Nilius and Goniakowski found the well-ordered ensembles of small-size distributed Cr and Fe atoms, on molybdenum-supported ultrathin magnesia with high temperature stability.[12] Pacchioni et al.[15] investigated the adsorption behavior of late transition metal atoms on metal supported magnesia (001), and found that the Cu, Ag and Au form full anions, exhibiting drastically different adsorption sites, bond distances and magnetic states compared with adsorption properties on bare magnesia (001). The EPR measurements confirmed that the adsorption sites of alkali adatoms on thin oxide film are closely influenced by atom type and experimental temperature.[16] Pacchioni et al. studied the charging mechanism of gold atoms and the stabilization of two dimensional charged species on defect-free thin magnesia film grown on molybdenum.[17, 18] The charging behavior of gold dimer and clusters on ultrathin magnesia (001) surface are confirmed by scanning probe techniques, and found that the gold dimer exists in both upright and flat lying geometries because of the flat potential surface.[19, 20] Freund et al.[21] reviewed microscopically the gold-oxide film interaction and its important impact on the physiochemical properties of adsorbed gold atoms and nanoparticles. In addition to the preferred strong adsorption of metal clusters, several important gas molecules and organic substances (such as molecular oxygen, dihydrogen, carbon dioxide, nitric oxide, methanol) can also be activated greatly on thin oxide films, induced by the

interplay between interfacial geometry and the electronic properties.[11, 22-26]

The generation and adsorption of high reactive oxygen species play essential role in heterogeneous catalysis processes and environmental chemical processes.[27, 28] Recently, Kim and Kawai et al. investigated deeply the dissociation behavior of a single water molecule on ultrathin MgO films, and found that the dissociative products are stabilized on MgO/Ag(001) due to the strong hybridization effect of electronic states at the oxide-metal interface.[29] By means of STM at cryogenic temperatures, Kim and Kawai et al.[30] proposed two different approaches for dissociating a single water molecule, namely excitation of vibrational states or excitation of electronic states. During the dissociation reaction initiated by the excitation of vibrational states, a hydroxyl group was produced and stabilized on the surface, while the dissociation product by the excitation of electronic states is atomic oxygen. Giordano and Ferrari[31] studied the modified ion pair interaction and obtained the noticeably stabilized hydroxyl groups on MgO ultrathin surface barrierlessly. By employing HREELS and XPS spectroscopy, Savio et al.[32] observed strongly enhanced probability of water dissociation on monolayer and submonolayer magnesia films, indicative of active role of Ag substrate and low coordinated ions at the border of monolayer magnesia islands. Using ambient pressure XPS, Bluhm et al.[33] observed an abrupt onset of hydroxylation near 0.01% relative humidity, due to water molecules aggregating at the surface. On supported magnesia, Grönbeck et al.[34] found the hydrogen bonded O-H stretching vibration are red shifted ~ 200 cm$^{-1}$, compared with counterparts on bare magnesia. For heavy loadings of water molecules, Sauer et al.[35] interpreted the reconstructed surface involving hydrated and hydroxylated magnesium ion using density functional theory combined with statistical thermodynamics. Further inspired by spectacular significant developments in the oxide-metal hybrid catalyst,[36, 37] our group have verified the hydroxyl groups, OOH species and the oxygen adatom during water dissociation, as well the coadsorption of water and oxygen on stoichiometric MgO ultrathin films.[38, 39] Previously, the variety of hydroxyl species obtained on MgO surfaces are derived from splitting water. The multiple methods to produce hydroxyl groups from other easily accessible hydroxide compounds are of important implications for controllable hydroxylation of oxide surfaces with special acid-base properties, and catalytic oxidation processes.

Hydrogen peroxide can be obtained from reaction between ozone and water, and is ubiquitous in terrestrial atmosphere (recognized to be a key component in the photochemistry of the earth's lower atmosphere).[40, 41] Significant fraction of hydrogen peroxide can be measured in precipitation and even in remote maritime regions.[41] However, hydrogen peroxide is widely regarded as a cytotoxic agent causing serious contact burn to human body and have high oxidative ability toward many organisms because of the metastable -1 valence state of oxygen. Under catalytic condition, hydrogen peroxide decomposes disproportionately to molecular oxygen and water. Thus hydrogen peroxide is regarded to be potential clean oxidant with the only reduction by-product water. The key aspects of hydrogen peroxide adsorption behavior on material surfaces are of significant importance in various areas such as environment, food, electrochemistry, biosensor and clinical laboratory.[42-46] Chemical interaction and decomposition of hydrogen peroxide taking place at fuel cells is an essential process in oxygen reduction reaction.[47, 48] Zhu et al. found the ultrathin $TiO_2$ nanosheet is an effective inductive agent for transferring $H_2O_2$ into reactive superoxide radicals.[49] Hydrogen peroxide can be fragmented by iron oxides to produce oxidative hydroxyl radicals in Fenton chemistry. Heterogeneous catalysts of iron oxide can enhance catalytic activity as compared with homogenous $Fe^{2+}$ catalysts.[50] The catalytic activity toward splitting hydrogen peroxide for hybrid catalyst, $Fe_3O_4$ combined with $Fe^0$ metal, is much higher than either its component oxide or metal system, which can be attributed to the thermodynamically favorable electron transfer between $Fe^0$ and $Fe_3O_4$.[51, 52] The metal oxides and hydrogen peroxide are common constituents of atmospheric and natural waters, and these two species frequently participate in oxidation processes used for treating a wide range of contaminants.[53, 54] As far as we know, energetically favorable dissociative adsorption state of single molecular hydrogen peroxide has never been obtained on perfect magnesia (001) surface.

In this contribution, for the first time the hydrogen peroxide and organic peroxides adsorption on the single crystalline MgO(001) films grown on transition metal substrates has been exploited systematically for enhancing the reactivity of insulating oxide surface toward generating hydroxyl species and reducing peroxides. In contrast with the extremely difficult fragmentation on bare magnesia, the dissociative adsorption energies of all the considered fragmentation configurations are substantially negative, demonstrating dissociation and reduction behavior on MgO(001)/TM are thermodynamically favorable. The structural and electronic properties are analyzed in detail to reveal the mechanism of reactivity enhancement for energetically and dynamically favorable decomposition of hydrogen peroxide and organic peroxides on metal-supported oxide film.

## 2. Models and methodologies

Periodic density functional theory (DFT-D2) calculations considering the long-range dispersion correction approach by Grimme[55] were performed applying the projector-augmented-wave (PAW) methods[56] implemented in the Vienna ab initio simulation package (VASP) code.[57, 58] All calculations are based on Kohn-Sham density functional theory with the generalized gradient approximation (GGA).[59] The exchange-correlation potentials are calculated using Perdew, Burke, and Ernzerhof (PBE) functional.[60] The sizes of plane-wave basis sets are determined by a large cutoff energy 500 eV to keep the accuracy of total energies. The valence orbitals are calculated by linear combining of plane waves. $(2\sqrt{2}\times2\sqrt{2})R45°$ MgO(001) surface supercells were used to eliminate interaction effect between periodic adsorbates. We used hybrid surface slabs consisting of 1 ML - 5 ML of MgO and 4 ML transition metal atoms, which can be denoted as 1 ML - 5 ML MgO(001)/TM(001). The geometric structures are optimized using $2 \times 2 \times 1$ gamma-centered k-point sampling of the Brillouin zone until atomic forces are less than 0.02 eV/Å. The two bottom layers of transition metal atoms are fixed to mimic the structural properties of bulk. 4 × 4 × 1 denser gamma-centered grids were used for electronic structure calculations. The amount of effective charge distribution is examined quantitatively using Bader program developed by Henkelman and coworkers.[61] The minimum energy pathways are investigated using the nudged elastic band method with climbing image modifications from the VTST code.[62] The VASPMO program,[63] Visual Molecular Dynamics (VMD) program,[64] and the VESTA program[65] are employed to analyze and visualize the obtained electronic and geometric structures.

## 3. Results and discussion

### 3.1 Hydrogen peroxide dissociation on pristine MgO

The obtained adsorption and dissociation structures of hydrogen peroxide on bare MgO (001) surface are shown in Figure 1. The nondissociative adsorption configuration is 0.99 eV lower in energy than the isolated molecular hydrogen peroxide and MgO (001) slab. The O-O bond length is 1.485 Å, slightly prolonged by the weak chemical adsorption of molecular hydrogen peroxide (O-O bond distance 1.471 Å). At the nondissociative adsorption state, the surface relaxation of the top layer MgO(001) is very small ($\Delta z$ = 0.098 Å), as listed in Table 1. The surface rumpling parameter is defined as the largest projection distance between surface oxygen and surface magnesium in $z$ direction. Owing to the strong electronegativity of oxygen, the O1H1 and O2H2 groups bind with the surface oxygen and form O-H···$O_s$ strong hydrogen bonds with short H···$O_s$ distances 1.579 Å and 1.572 Å, which should play major role in determining adsorption strength of hydrogen peroxide on the pristine MgO (001) surface. The approaching process from vacuum to surface may lead to the transformation of hydrogen peroxide to its mirror structure to obtain appropriate collision direction. Hydrogen peroxide transformation pathways to its mirror structure in vacuum and on pristine MgO(001) surface are compared in Figure 2. Although the reaction in vacuum exhibits an energy barrier 0.047 eV, the transformation reaction on surface is barrierless to produce the mirror isomer of hydrogen peroxide. After dissociation, hydroxyl groups bind with two neighboring surface magnesiums substantially more firmly than that of nondissociative adsorption state. The O1-O2 distance is lengthened to 3.365 Å and the dihedral angle is further decreased to 85.6, which demonstrate the equilibrium geometries of molecular hydrogen peroxide is thoroughly broken. The surface relaxation is more serious with rumpling value of 0.538 Å. The dissociative adsorption state is energetically unfavorable by 1.76 eV, comparing with the nondissociatvie adsorption structure. The transition state (as shown in Figure 3a and Table 1) has a high energy structure, with each hydroxyl group binding to one surface magnesium. The O-H, and O1-O2 distances and surface rumpling value lie between the nondissociative adsorption state and dissociative adsorption state. As the singly coordinated hydroxyl is much looser (single bonded hydroxyl can rotate on the surface more easily), the dihedral angle of H1O1O2H2 deviates significantly from the reactants and dissociation products. As show in Figure 3b, the dissociation of hydrogen peroxide is calculated to be highly endothermic process and the activation barrier is 1.85 eV, indicating it is extraordinarily difficult to dissociate hydrogen peroxide on pristine MgO(001) surface.

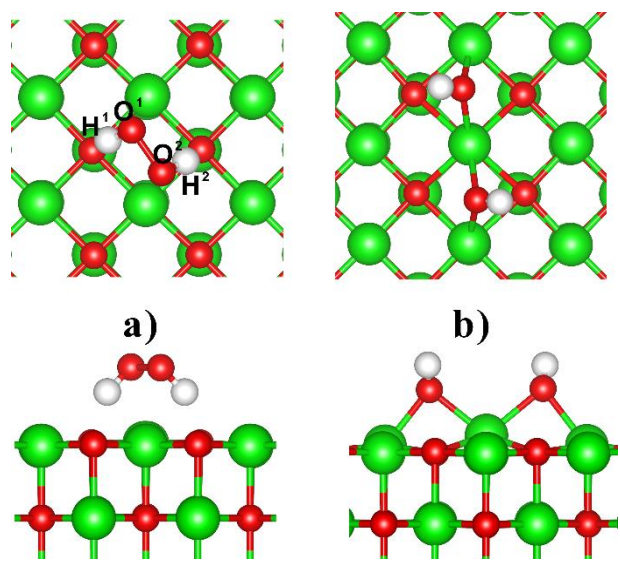

**Figure 1.** The atomic structures of hydrogen peroxide (a) nondissociative adsorption and (b) dissociative adsorption on pristine MgO(001) surface.

**Table 1.** The optimized structural parameters (bond length and surface rumpling $\Delta z$ in Å, dihedral angles $D$ of H1O1O2H2 in degree) of nondissociative adsorption state (A),

transition state (TS), and dissociative adsorption state (D) of hydrogen peroxide on bare MgO(001) surface.

| State | d(O1-O2) | d(O1-H1) | d(O2-H2) | d(O1-Mg) | d(O2-Mg) | D | Δz |
|---|---|---|---|---|---|---|---|
| A | 1.485 | 1.037 | 1.037 | 2.412 | 2.452 | 92.3 | 0.098 |
| D | 3.365 | 0.973 | 0.973 | 2.170, 2.134 | 2.161, 2.128 | 85.6 | 0.538 |
| TS | 3.340 | 0.974 | 0.977 | 2.874, 2.026 | 2.647, 2.037 | 121.8 | 0.271 |

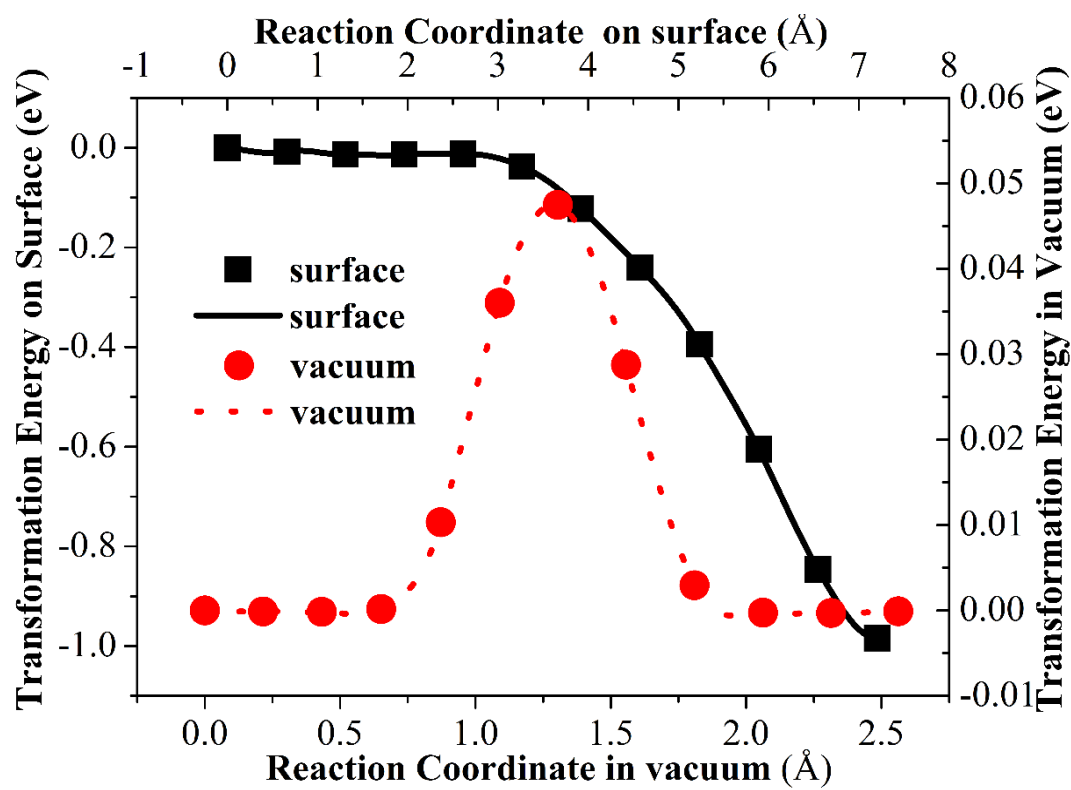

**Figure 2**. The reaction energy profile for hydrogen peroxide transformation to its mirror structure in vacuum and on pristine MgO(001) surface. For reaction on pristine MgO(001) surface, in initial structure, the hydrogen peroxide is placed in the vacuum layer 5 Å away from the surface.

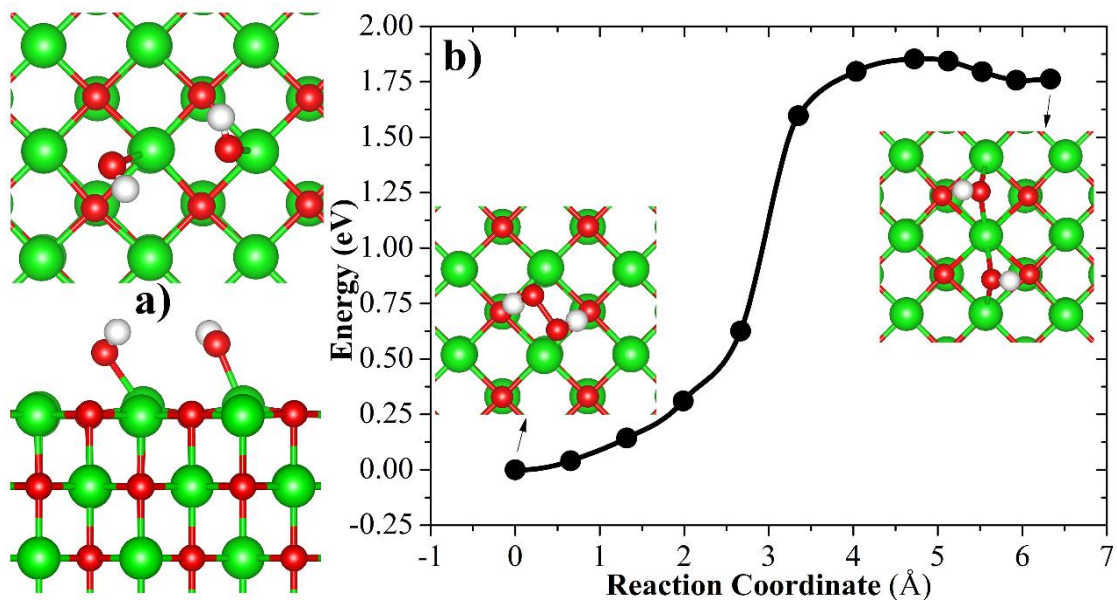

**Figure 3.** The transition state structure (a) and reaction energy profile (b) for hydrogen peroxide dissociation on pristine MgO(001) surface.

## 3.2 Hydrogen peroxide dissociation on MgO(001)/Mo(001) surfaces

In order to increase the reactivity of insulating MgO(001) surface, we investigated the adsorption behaviors and dynamics of molecular hydrogen peroxide on MgO(001) films supported on Mo(001) single crystal. Strikingly, as shown in Figure 4, the hydrogen peroxide molecules were completely dissociated to surface hydroxyl groups on MgO(001)/Mo(001). The structural parameters for hydrogen peroxide dissociative adsorption on 1 ML - 10 ML MgO(001)/Mo(001) are shown in Table 2. The $O^1$-$O^2$ distances are lengthened to larger than 3.4 Å (3.401, 3.447, 3.434, 3.435, 3.426, 3.457 Å for 1-5 and 10 ML systems, respectively), indicating the complete fragmentation of peroxide bonds. The O1-O2 distances show slight even-odd alteration, and the O1-O2 distances on oxide films with even-numbered layers are larger than that on oxides with neighboring odd-numbered layers. The dihedral angles of bond H1O1O2H2 (calculated to be 112.9° at molecular ground state) are decreased to 87.5° - 89.3° for 1 ML – 3 ML systems. The 4 ML, 5 ML and 10 ML systems show much larger dihedral angles (94.0°, 95.9°, and 103.6°, respectively). Compared with the molecular adsorbed structure on bare magnesia surface, the O1-H1 and O2-H2 bond lengths of dissociative hydrogen peroxide on metal are reduced to ca. 0.97 Å, which suggest that the interaction between two OH groups are weak and surface hydroxyls with highly ionic bonding components formed after dissociation. Mg1-O1 (range from 1.959 Å to 1.980 Å for 1 ML – 10 ML systems) and Mg3-O2 (range from 1.939 Å to 1.987 Å for 1 ML – 10 ML systems) distances are prolonged with increasing film thicknesses. Mg2-O1 and Mg2-O2 distances of systems with even-numbered layers show larger distances than that of neighboring systems with odd-numbered layers. The O1 and O2 both bind with two surface magnesium firmly, with bond distances 1.94 Å – 1.99 Å, which imply the much stronger interaction between dissociative hydrogen peroxide and the metal supported surface, compared with the case of pristine MgO(001) surface. The Mg1-$O_s$1, Mg1-$O_s$2, Mg2-$O_s$3, Mg2-$O_s$4 are destroyed, showing corresponding Mg-O distances ranging from 3.43 Å to 3.60 Å. The $O_s$7, $O_s$8, $O_s$9 are the second layer oxygen atoms located beneath the surface magnesium Mg1, Mg2, Mg3. Mg1-$O_s$7 and Mg2-$O_s$8 distances range from 2.97 Å – 3.43 Å, while the Mg3-$O_s$9 have much shorter bond length about 2.2 Å. For reactions on 1 ML - 10 ML MgO(001)/Mo(001) films, surface ionic bonds experienced different levels of relaxation and rupture. As a consequence, after dissociation reaction, the five-coordinated surface magnesium atoms possess even lower coordination number. For dissociation reaction on 1 ML MgO(00)/Mo(001) film, the coordination numbers for Mg1, Mg2, and Mg3 are three, four, three respectively. For reaction on 2 - 3 ML MgO(001)/Mo(001) films, the coordination numbers for Mg1, Mg2, and Mg3 are three, four, four respectively, because of the nonbreaking Mg3-$O_s$9 bonds. Apparently, the surface rumpling (1.10 Å – 1.19 Å) are more remarkable than that of pristine oxide film. The maximum and minimum bond distances in *z* direction for hydrogen peroxide fragmentation on bulk magnesia (001) and molybdenum-supported 1 ML, 2 ML, 3 ML and 10 ML (f) magnesia (001) films are depicted in Figure S1. Before adsorption of hydrogen peroxide, metal-supported surface show very small difference among Mo-O distances. For hydrogen peroxide fragmentation on 1 ML – 2 ML magnesia films, the Mo-O bonds experience large distortion, with largest Mo-O distance difference 0.158 Å and 0.113 Å, respectively. For thicker films, the Mo-O distortion becomes very small, with largest Mo-O distance difference 0.029 Å and

0.005 Å for 2 ML and 10 ML oxide films, respectively. The Mo-O bonds of thick films approach those before adsorption of hydrogen peroxide. For the 2 ML, 3 ML and 10 ML oxide films, the (O-Mg)$_z$ bonds of the first oxide layer exhibit significant distortion, with largest bond distances 1.325 Å, 1.43 Å and 1.3 Å, respectively, indicating that the surface layer oxide undergoes structural mutation to accommodate the produced hydroxyls groups. For the inner (O-Mg)$_z$ bonds, the difference between maximum and minimum of bond distance is much smaller. After dissociation of hydrogen peroxide on 3 ML and 10 ML oxide films, the (O-Mg)$_z$ bonds of second layer show bond differences 0.188 Å and 0.182 Å, respectively. Moreover, the dissociative adsorption of hydrogen peroxide also leads to the severe structural deformation of (O-Mg)$_z$ bonds, while the molecular adsorption of hydrogen peroxide on pristine magnesia (001) affects the surface structure only slightly.

**Table 2.** The optimized structural parameters (bond length and surface rumpling $\Delta z$ in Å, dihedral angles in degree) of dissociative adsorption state of hydrogen peroxide on 1 ML – 5 ML MgO(001)/Mo(001) surfaces.

|  | 1 ML | 2 ML | 3 ML | 4 ML | 5 ML | 10 ML |
| --- | --- | --- | --- | --- | --- | --- |
| d(O1-O2) | 3.401 | 3.447 | 3.434 | 3.435 | 3.426 | 3.457 |
| d(O1-H1) | 0.969 | 0.968 | 0.968 | 0.969 | 0.969 | 0.969 |
| d(O2-H2) | 0.967 | 0.968 | 0.968 | 0.968 | 0.968 | 0.970 |
| D(H1O1O2H2) | 87.5 | 87.9 | 89.3 | 94.0 | 95.9 | 103.6 |
| d(Mg1-O1) | 1.959 | 1.962 | 1.963 | 1.967 | 1.970 | 1.980 |
| d(Mg1-O$_s$1) | 3.603 | 3.552 | 3.517 | 3.476 | 3.462 | 3.443 |
| d(Mg1-O$_s$2) | 3.439 | 3.562 | 3.510 | 3.474 | 3.469 | 3.430 |
| d(Mg1-O$_s$3) | 1.981 | 1.954 | 1.953 | 1.953 | 1.954 | 1.951 |
| d(Mg1-O$_s$4) | 1.976 | 1.957 | 1.958 | 1.959 | 1.962 | 1.959 |
| d(Mg1-O$_s$7) | — | 2.972 | 3.061 | 2.976 | 2.973 | 2.898 |
| d(Mg2-O1) | 1.953 | 1.968 | 1.968 | 1.973 | 1.971 | 1.978 |
| d(Mg2-O2) | 1.962 | 1.966 | 1.961 | 1.965 | 1.965 | 1.971 |
| d(Mg2-$O_s$3) | 3.586 | 3.572 | 3.519 | 3.494 | 3.480 | 3.483 |
| d(Mg2-$O_s$4) | 3.565 | 3.579 | 3.548 | 3.525 | 3.535 | 3.511 |
| d(Mg2-$O_s$5) | 2.097 | 2.065 | 2.065 | 2.062 | 2.060 | 2.065 |
| d(Mg2-$O_s$6) | 2.094 | 2.077 | 2.084 | 2.081 | 2.084 | 2.081 |
| d(Mg2-$O_s$8) | — | 3.329 | 3.432 | 3.374 | 3.364 | 3.333 |
| d(Mg3-O2) | 1.939 | 1.972 | 1.973 | 1.978 | 1.982 | 1.987 |
| d(Mg3-$O_s$5) | 3.075 | 2.879 | 2.837 | 2.818 | 2.814 | 2.814 |
| d(Mg3-$O_s$6) | 2.953 | 2.778 | 2.736 | 2.716 | 2.708 | 2.711 |
| d(Mg3-$O_s$9) | — | 2.201 | 2.258 | 2.261 | 2.261 | 2.199 |
| d(Mg3-$O_s$10) | 2.018 | 2.013 | 2.021 | 2.023 | 2.025 | 2.033 |
| d(Mg3-$O_s$11) | 1.985 | 1.991 | 1.999 | 2.001 | 2.002 | 2.009 |
| $\Delta z$ | 1.190 | 1.138 | 1.136 | 1.107 | 1.105 | 1.104 |

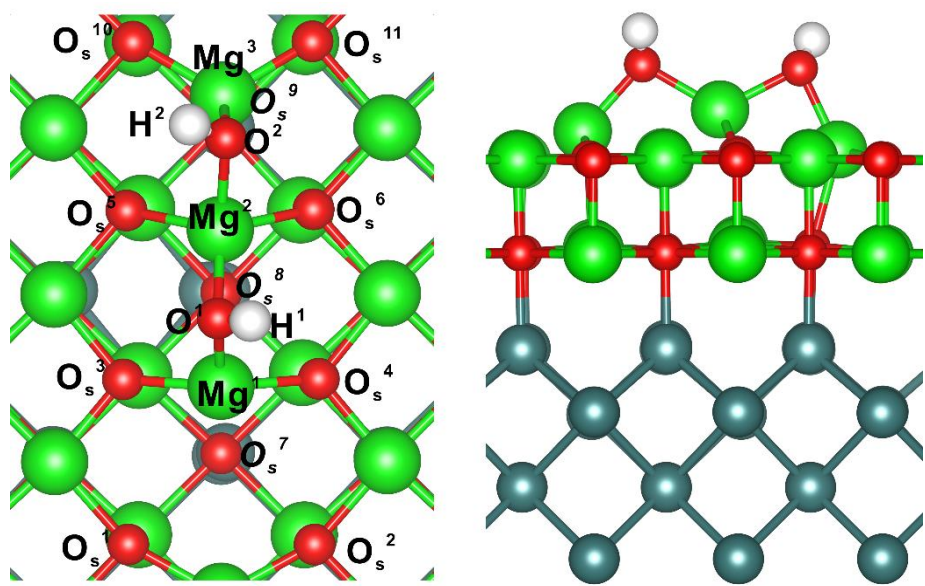

**Figure 4.** Hydrogen peroxide dissociation on MgO(001)/Mo(001). Top (left) and side (right) view of the optimized configuration of hydrogen peroxide dissociative adsorption on 2 ML MgO(001)/Mo(001).

The theoretical models for similar hybrid system always contain oxide films no more than 5 ML and we have also adopted the idea of the theoretical model (namely the ultrathin oxide films are used).[17, 18] As the stress induced by lattice mismatch between perfect oxide film and metallic support becomes very large with increasing film thickness, the line defects and dislocation networks appear extensively in the thick oxide film to compensate the lattice mismatch and relax the strain force.[66] The STM (scanning tunneling microscopy) experiments revealed the electronic structure of thin oxide films with various thicknesses and found that the magnesia film with thickness larger than 5 ML approach properties of bulk material.[67-70] Thus, we only consider the adsorption energies of hydrogen peroxide on 1 ML – 10 ML ultrathin magnesia films here. The dissociative adsorption energies are calculated to be -6.82 eV, -6.10 eV, -5.36 eV, -4.64 eV and -4.02 eV for hydrogen peroxide dissociation reaction on 1 ML – 5 ML MgO(001)/Mo(001) surfaces. To adequately reveal the influence of film thickness on reaction thermodynamics, we present the adsorption energy and total energy *vs* oxide thickness as shown in Figure 5 and Table S1. As the surface structure characteristics are similar for hydrogen peroxide dissociation on 1 ML – 10 ML films, the total energy rises linearly with increasing film thickness. The total energy increase is nearly the energy of subsequently added oxide layers. The dissociative adsorption energy decreases monotonously with increasing film thickness, demonstrating the lower reactive activity of thick oxide films. The energy profiles are illustrated to reveal the reaction dynamics during the hydrogen peroxide splitting, as shown in Figure 6. The larger reaction coordinate of dissociation pathway on thinner films can be attributed to the more serious surface relaxation upon dissociation and formation of high oxidizing hydroxyls. Obviously, the hydrogen peroxide molecule can be dissociated smoothly without any activation barrier, on metal supported ultrathin oxide films.

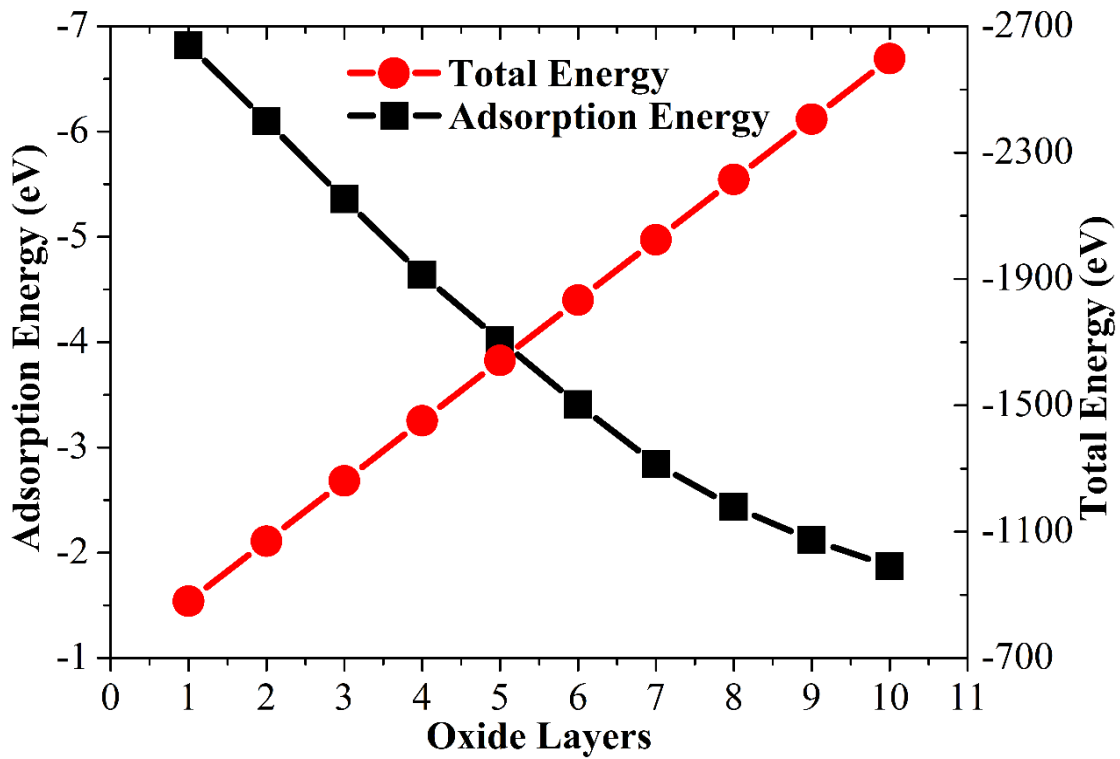

**Figure 5.** The adsorption energies and total energies *vs* oxide thickness, for dissociative hydrogen peroxide adsorbing on 1 ML – 10 ML MgO(001)/Mo(001).

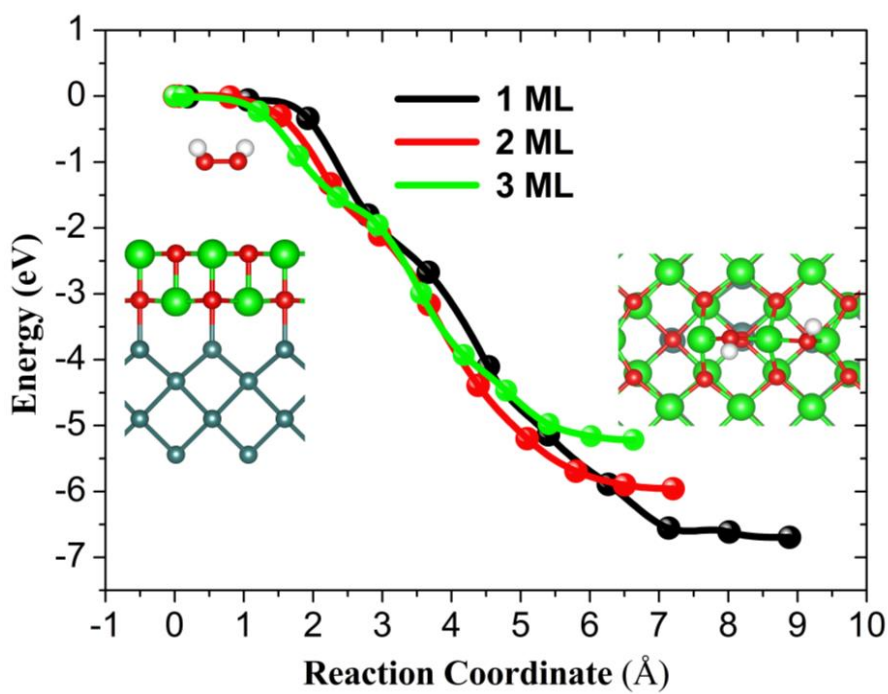

**Figure 6.** Reaction energy profile for dissociation reaction of hydrogen peroxide on MgO(001)/Mo(001) films with different thicknesses.

After hydrogen peroxide adsorption, the initiative larger surface rumpling of magnesia film deposited on molybdenum can further activate the surface, which can be part of the reason for facilitating hydrogen peroxide dissociation on MgO/Mo hybrid surface. To compare with the hydrogen peroxide dissociation on bare magnesia (001) with small surface rumpling, we introduce a model with all atoms of hybrid surface fixed and then relax the adsorbate (starting from the molecular hydrogen peroxide). We find that the hydrogen peroxide could be dissociated spontaneously, releasing heat of -2.56 eV, without distortion of surface Mg-O bonds, as shown in Figures 7a. When deposited on the molybdenum, the lattice of magnesia is extended by 5%. Could peroxide bonding be destroyed by bare extended magnesia (001) surface? We introduce another model with bare magnesia (001) surface extended to same size as MgO/Mo hybrid surface to investigate the hydrogen peroxide adsorption behavior. Indeed, the adsorption structure shows different feature compared with that on non-extended magnesia surface (as shown in Figures 7c and 7d). The extended surface split the O-H bonds of hydrogen peroxide, with dissociative adsorption energy of -1.78 eV. However, the dissociative state with broken peroxide bond is 1.10 eV higher in energy than dissociative state with broken O-H bonds. Thus, without the metal substrate, the dissociative state with broken O-H bonds is much more preferred than molecular adsorption state and dissociative state with broken peroxide bond on extended bare magnesia surface (5%). From the comparison between the dissociation behavior on bare magnesia, extended bare magnesia, and metal-supported magnesia, it can be deduced that the effect of metal substrate should play an important role in dissociation and reduction of hydrogen peroxide.

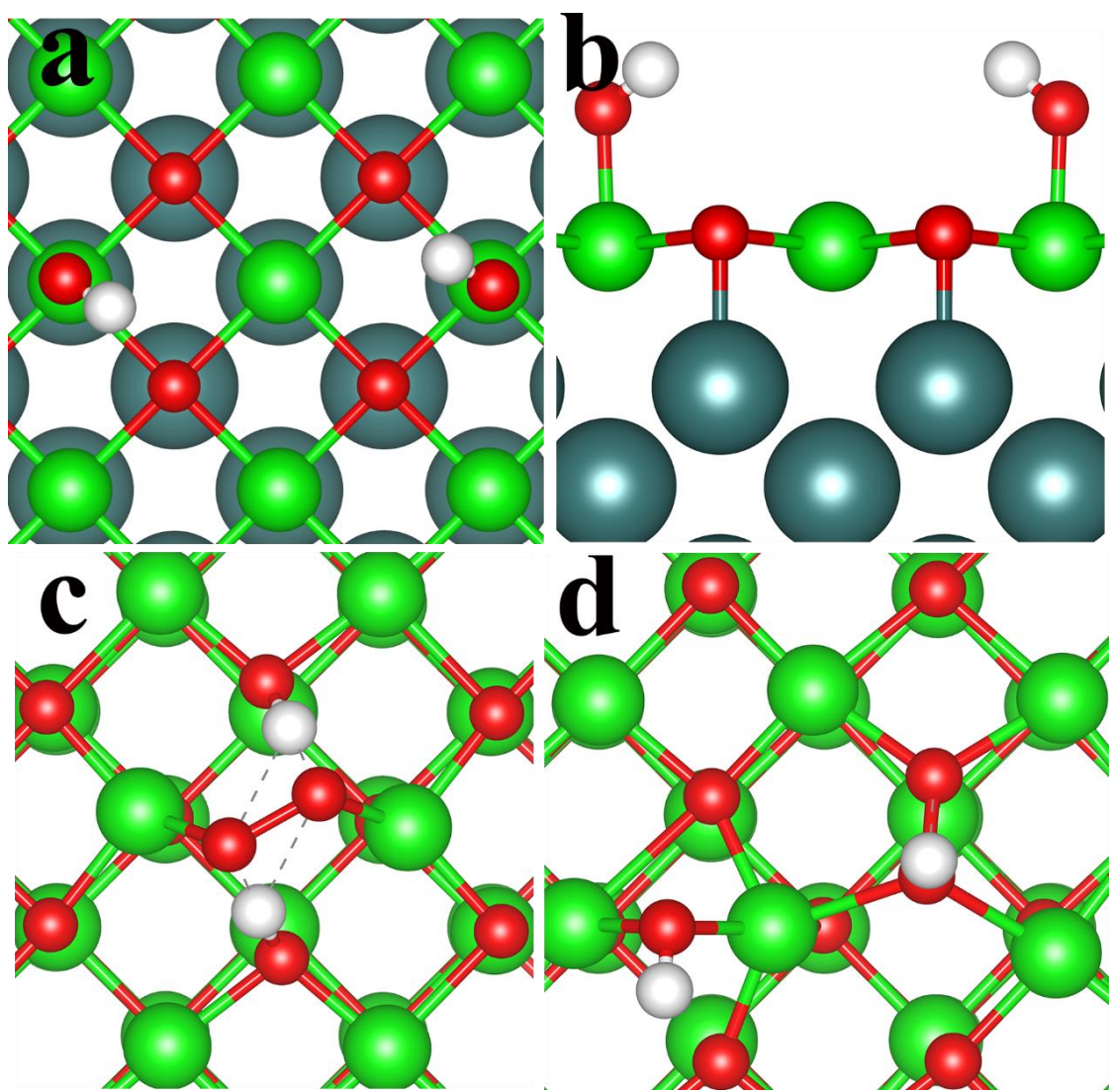

**Figure 7.** Top (a) and side (b) views of the hydrogen peroxide dissociation on MgO(001)/Mo(001) without surface distortion. The dissociative state with broken

O-H bonds (c) and broken peroxide bond (d) on extended bare magnesia surface with cell parameters of molybdenum substrate.

### 3.3 Characterization and discussion of electronic properties

The phenomenon and mechanism of reactivity enhancement on ultrathin oxide films (1 ML - 3 ML) can be further understood by characterizing the electronic properties of interface structures. As listed in Table 3, in contrast with the zero total valence of OH in molecular hydrogen peroxide, the O1-H1 and O2-H2 groups on MgO(001)/Mo(001) surfaces show negative charges of -0.820 e ~ -0.831 e, verifying the formation of hydroxyl species and the strong chemical bonding interaction between adsorbates and the surfaces. Whether before the adsorption or after the dissociative adsorption of hydrogen peroxide, the charge values of surface oxygen increase with thickness of oxide film, suggesting that the surface oxygen of thicker oxide films is less affected by the substrate and retain relatively high ionic character comparing with the thinner oxide films. After the adsorption of hydrogen peroxide, the net charge of surface oxygen decreases, indicating the formation of oxidizing groups with high electron affinity on the metal-supported MgO (001) and bare MgO (001). The surface oxygen of 1 ML MgO (001) carries least charge, indicating the electronic property of metal-supported one monolayer magnesia are significantly different from the thick films and pristine magnesia (001). The hydroxyl species on MgO are also negatively charged, while the charge values of hydroxyl are much smaller (-0.54 e and -0.53 e), indicating the weaker binding on bare MgO (001). Because of the electron withdrawing effect of high oxidizing hydroxyl groups, the low coordinated magnesium atoms Mg1, Mg2, Mg3 present almost the same valence state with other five coordinated surface magnesium (Mg($1^{st}$ L)). The Mo substrate show negative charges, due to high electron affinity of molybdenum.

**Table 3.** Charge distributions of dissociated hydrogen peroxide (O1H1 and O2H2 groups), surface magnesium at the reaction site, surface oxygen, based on Bader charge analysis (unit in electron).

| Species | 1 ML MgO/Mo | 2 ML MgO/Mo | 3 ML MgO/Mo | MgO |
|---------|-------------|-------------|-------------|-----|
| O1H1    | -0.820 | -0.828 | -0.828 | -0.540 |
| O2H2    | -0.819 | -0.831 | -0.830 | -0.533 |
| Mg1     | +1.641 | +1.655 | +1.655 | +1.675 |
| Mg2     | +1.650 | +1.652 | +1.653 | +1.695 |
| Mg3     | +1.617 | +1.651 | +1.652 | +1.676 |
| $O_{s0}$[a] | -1.519 | -1.656 | -1.658 | -1.653 |
| $O_{s1}$[a] | -1.485 | -1.635 | -1.641 | -1.601 |
| Mg(1st L) | +1.643 | +1.663 | +1.663 | +1.655 |
| Mo      | -0.731 | -0.432 | -0.527 | —      |

[a] The charge of $O_{s0}$ and $O_{s1}$ are averaged over the surface oxygen before and after the dissociative adsorption of hydrogen peroxide, respectively.

The dissociative state of hydrogen peroxide can be proved by examining the localized and projected density of states of adsorbates, surface, and interface, as shown in Figure 8. The O1H1 and O2H2 groups have similar peak character. However, in energy levels they do not superpose with each other as in the molecular hydrogen peroxide, due to different chemical circumstances of the two hydroxyl groups. The electronic states of O2H2 tend to occupy higher energy levels comparing with $O^1H^1$, which imply the O2H2 possess higher chemical activity than O1H1. At energy levels around -8 eV, the Mg1 and Mg3 states coincides with the O1H1 and O2H2 respectively. In addition, the states of Mg2 hybridizes with both O1H1 and O2H2 around -8 eV. However, other surface magnesium atoms do not show any occupation number at this energy level (around -8 eV). The Mg1, Mg2, Mg3 all show characteristic peaks at -4.9 eV, which overlap with both O1H1 and O2H2. These electronic structural evidences further demonstrate that the Mg1 and Mg3 forms strong chemical bonds with O1 and O2 respectively, while the Mg2 binds chemically to both O1 and O2. Consequently, at the energy levels close to Fermi level (around -2.6 eV), the $Mg^{1st\,L}$ atoms possess more high energy electrons and the Mg2 are most stabilized. Although the differences in electron affinity and chemical circumstances of oxygen and molybdenum lead to different electron density values, the electron density plots of *z* orientation character of Mo-4d and O-2p orbitals show very similar and synchronous ups and downs in large energy ranges -6.6 eV ~ -2.7 eV, indicating the strong orbital hybridization and covalent bonds formation between interfacial oxygen and molybdenum.

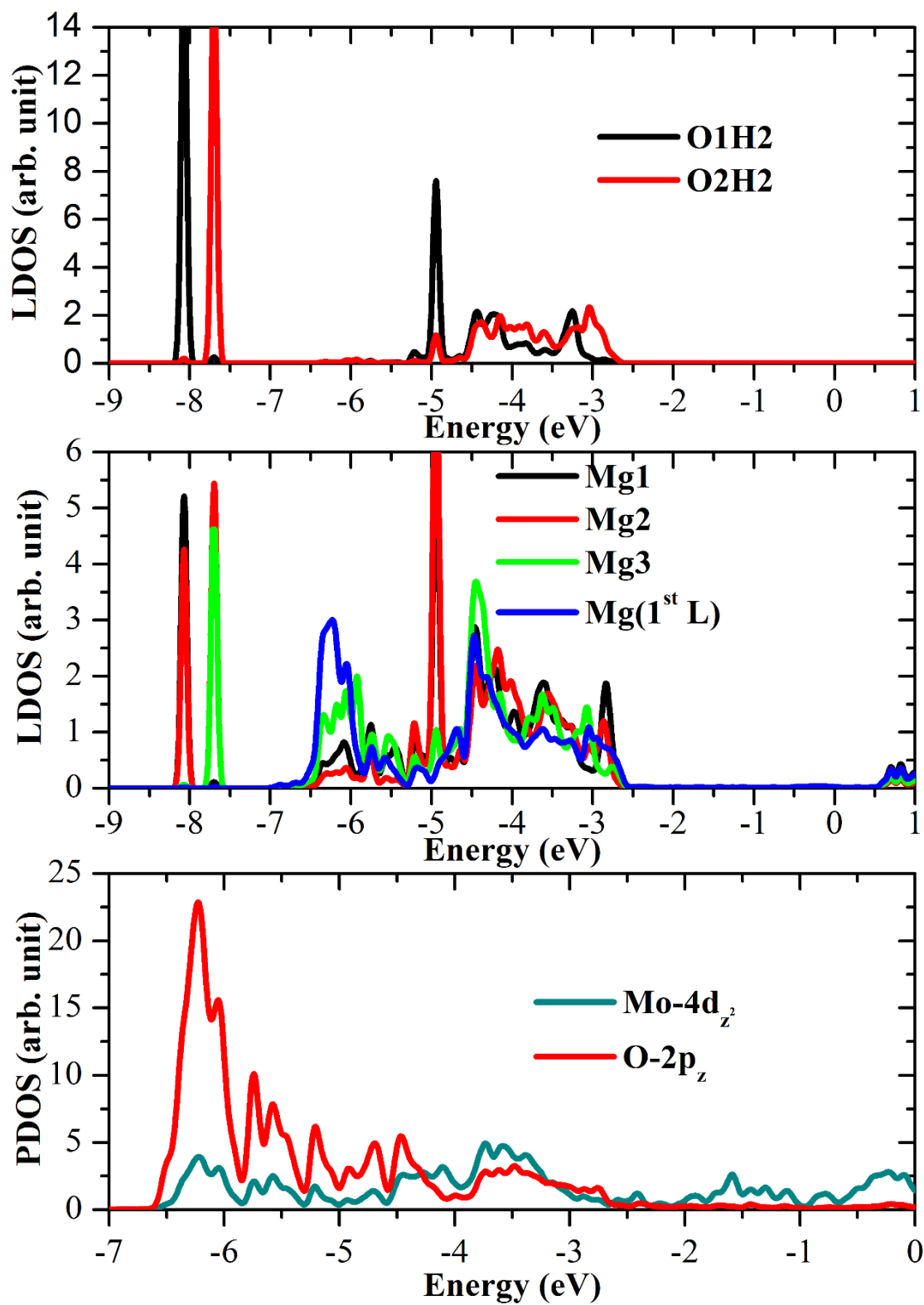

**Figure 8.** Density of states analysis of dissociative hydrogen peroxide on 2 ML MgO(001)/Mo(001): Localized density of states (LDOS) of hydroxyl groups O1H1 and O2H2 (top); LDOS of surface magnesium Mg1, Mg2, Mg3, and Mg(1st L) (middle); Projected density of states (PDOS) in $z$ orientation of Mo-4d and O-2p

orbitals in interfacial molybdenum and oxygen (bottom). The density of states of Mg1, Mg2 and Mg3 are multiplied by thirteen times for appropriate comparison with that of other surface magnesium atoms. The inset chart in middle figure is amplified to show the density of states between -2.7 eV ~ -2.5 eV. The Fermi energy level is set to be zero.

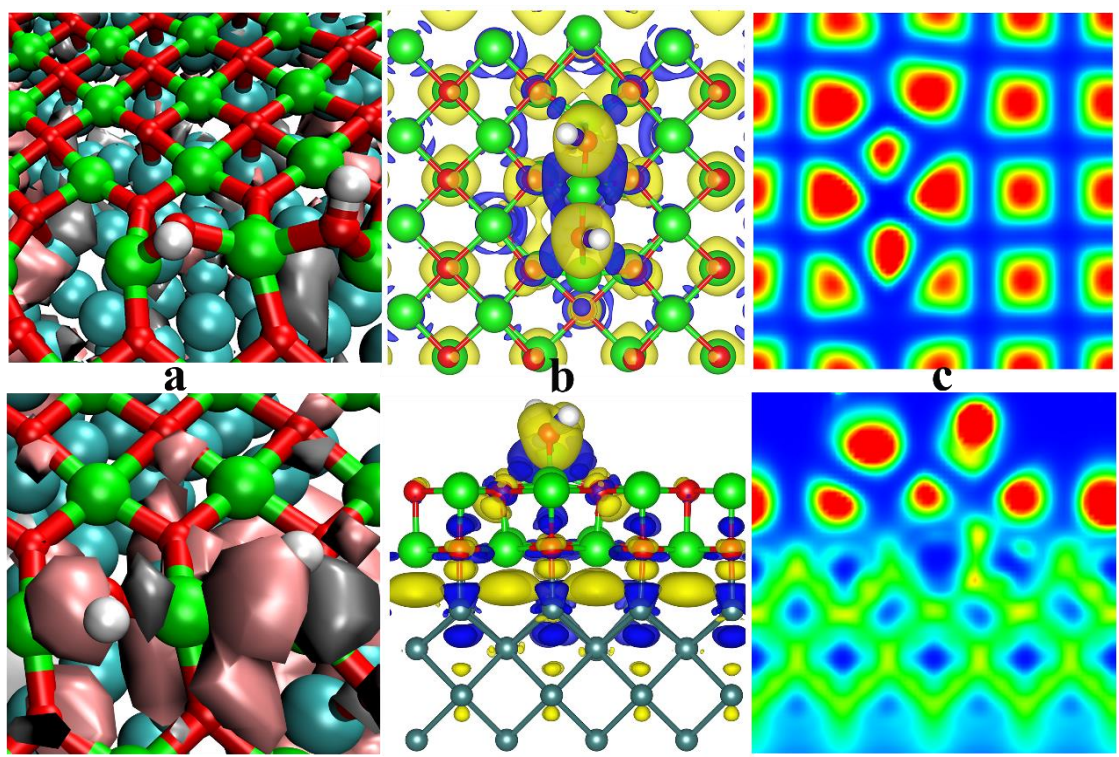

**Figure 9.** (a) The highest occupied orbital (top) and occupied orbital with largest $p_y$ coefficients of hydroxyl (bottom) for hydrogen peroxide dissociation on MgO(001)/Mo(001). The silver and pink colors represent negative and positive orbital phases, respectively. The absolute value of isosurface for the orbital is 0.01 a. u. (b) Top view (top) and side view (bottom) of the differential charge density contour in unit of e/bohr$^3$ for hydrogen peroxide dissociation on MgO(001)/Mo(001). The differential charge density is obtained by subtracting charge density of the adsorbed hydroxyl groups, the MgO film and the molybdenum substrate from the whole system. The isosurface value is set to be 0.003 e bohr$^{-3}$. The blue and yellow colors stand for the electron loss and electron gain, respectively. (c) The top view (top) and side view (bottom) of electron localization function for hydrogen peroxide decomposition on MgO(001)/Mo(001). The blue and red colors represent the electron delocalization and electron localization, respectively.

The highest occupied orbital, as shown in Figure 9a (top), mainly distributes in the molybdenum substrate and a few slices of highest occupied orbital spread to the interfacial area. This result indicates that after deposited on metallic substrate, the magnesia itself doesn't show obvious electron distribution at fermi level, which agrees well with the density of states of interfacial oxygen. The differential charge density contour for hydrogen peroxide dissociation on 2 ML MgO(001)/Mo(001) is calculated by density functional methods and illustrated in Figure 9b. Due to significantly different bonding environment and coordination number, Mg1, Mg2, Mg3 have different extent of charge depletion. The electron loss for Mg2 is most obvious. This is in accordance with the density of states analysis shown in Figure 8 (middle). As can be vividly seen in Figure 9b (bottom), at the interfacial area, oxygen and molybdenum form chemical bonds with electron accumulation in oxygen and electron depletion in molybdenum. Large areas of charge accumulation right under magnesium atoms can be seen in the interface, which can be attributed to bonding effect between interfacial oxygen and molybdenum and the weakening of the interfacial ionic bonds of MgO film after deposited on metal substrate. We analyzed the electron localization function (as depict it in Figure 9c), because the electron localization are essential for describing where local groups of electrons, electron pairs and unpaired electrons are placed.[71] After deposited on metallic slab, the oxygen ions show obvious electron localization effect and O-Mg bonding in magnesia remains highly ionic. In addition, the hydroxyl, which is decomposition product of hydrogen peroxide, also exhibits significant electron localization effect, indicating the formation of ionic bonds between adsorbates and surface magnesium. The molybdenum atoms linked to the oxygen of magnesia mainly exhibit electron delocalization effect and the electrons grasped from the magnesia are distributed extensively in the molybdenum substrate with high electron affinity, rather than localized at a few interfacial molybdenum atoms.

**3.5 Hydrogen peroxide dissociation on MgO(001)/TM(001)**

We have examined the suitability of other transition metals (silver, vanadium, tungsten and gold) supported insulating oxide for splitting hydrogen peroxide. Similar to the calculation of hydrogen peroxide adsorption on molybdenum supported magnesia (001), the bottom two layers of metallic substrate are fixed to mimic the properties of bulk transition metals. The cell size of MgO/TM hybrid surface are determined by the lattice constants of substrate, and the magnesia films are deposited on the metallic substrate. Compared with the bare MgO (001) surface, the lattice constants of gold, silver and vanadium slabs are 2.0%, 1.8% and 0.6% shrunk, respectively. Contrastively, the lattice constants of molybdenum and tungsten are 5.1% and 5.8% extended, respectively. The dissociative adsorption energies of hydrogen peroxide on MgO(001)/TM(001) are correlated to the lattice constants of substrate slabs, as shown in Figure 10. All the considered transition metal substrates supported MgO films can dissociate hydrogen peroxide to hydroxyls. The charge transfer from the surfaces to dissociative adsorbates are calculated to be around 1.7 e. The absolute value of dissociative adsorption energies are much larger when the lattice constants of substrate slabs increase. The extensive exploration confirm that the metal supported oxide surfaces in this study provide appropriate versatile model for understanding the dissociation and reduction behavior of peroxide bond.

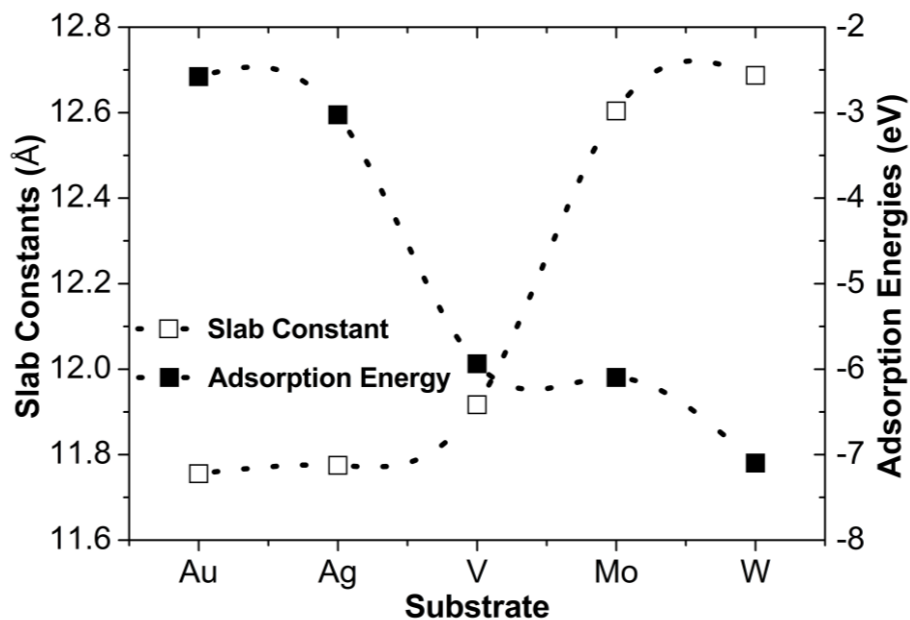

**Figure 10.** The dissociative adsorption energies of hydrogen peroxide on 2 ML MgO(001)/TM(001) (TM = Au, Ag, V, Mo, W) correlated to the lattice constants of substrate slabs.

## 3.6 The translation of hydroxyl group of hydrogen peroxide at dissociative adsorption state

The hydroxyl group of water molecule can translate on metal supported ultrathin oxide films, which have been studied by low-temperature scanning tunneling microscopy and density functional theory calculations.[30] In this respect, here we examine the further translation of hydroxyl of hydrogen peroxide at the dissociative adsorption state, and various dissociative adsorption structures, as shown in Figure 11 and Table 4. The ground state with largest dissociative adsorption energy of -6.82 eV is structure **11a** (Figure 11a). Dissociative adsorption of hydrogen peroxide on three neighboring magnesium forming rectangular angles produces structure **11b**, with adsorption energy of 6.403 eV. Structure **11c** (with adsorption energy of 6.402) is nearly isoenergetic with 9b. In structure **11c**, the two hydroxyl groups point toward the same direction and form hydrogen bonds with neighboring surface oxygen. Hydroxyl translating apart from each other leads to the formation of structure **11d**, with adsorption energy of 6.24 eV. Similar to structure **11c**, structure **11e** possesses two hydroxyl groups pointing toward the same direction, which deviate much the dihedral angle of molecular hydrogen peroxide. The dissociative adsorption energy of structure **11e** releases nearly the same amount of energy (6.24 eV) with **11d**. Comparing **11e** with **11d**, the very large change of the pointing direction of hydroxyl does not raise the total energy considerably, due to the retaining of surface structure characteristics. Dissociating hydrogen peroxide on two opposite sides of magnesium tetragon (possessing four neighboring magnesium) yields the structure **11f**, calculated to be 0.64 eV higher in energy than **11a**. Moving hydroxyl of **11a** to bind with the diagonally adjacent magnesium of Mg2 produces the configuration **11g**, with adsorption energy of -0.67 eV. Further moving hydroxyl of **11g** to bind with diagonally adjacent magnesium of Mg3 produces the configurations **11h and 11i**, with adsorption energies of -5.89 eV and -5.81 eV, respectively. Revolving a hydroxyl of structure **11f** toward opposite direction produces **11j** with much smaller dissociative adsorption energy -5.29 eV. This fact suggests that the hydrogen bond in **11f** is substantially strong and there is a cost of energy for breaking the hydrogen bonding interaction. Besides, it can be observed that one surface oxygen in **11f** forms two covalent Mo-O bonds (2.017 Å and 2.118 Å), which also contributes to total energy reduction. Splitting hydrogen peroxide to water and oxygen adatom yields the configuration **11k**, with dissociative adsorption energy of -4.72 eV. It can be seen the dissociative water is obtained in **11k**, with strong hydrogen bond between HO$_{water}$ and the newly formed surface hydroxyl (with hydrogen bond distance 1.548 Å). The hydrogen peroxide can be dissociated to OOH group, surface hydride (structure **11l**, with adsorption energy of -2.19 eV) or surface hydroxyl (structures **11m** and **11n**, with adsorption energies of -2 eV and -1.10 eV, respectively). The hydride ion in **11l** binds with two surface magnesium atoms, with H-Mg bond distances of 1.821 Å and 1.853 Å. The hydrogen bond in **11l** with distance of 1.479 Å should be substantially strong. For structure **11m**, the hydrogen and terminal oxygen in OOH group form strong hydrogen bonds with surface oxygen (O$_s$) and the newly formed surface hydroxyl (O$_s$H), with bond distances 1.338 Å (OOH...O$_s$) and 1.360 Å (O$_s$H...OOH), respectively. However, the corresponding hydrogen bonds are destroyed in structure

**11n**, which lead to relative energy raise by 0.90 eV. The structure **11o** with small dissociative adsorption energy of -0.85 eV, releases dihydrogen and forms superoxygen species on the surface. As can be seen from the relative adsorption energies shown in Table 4, the configuration **11a**, which has been discussed in detail, should play dominant role in the dissociation process. Drastically different from the dissociation structure on bare magnesia (001) with positive adsorption energy 0.77 eV, the dissociative adsorption energies of all the considered fragmentation configurations in Figure 11 are substantially negative, demonstrating that dissociation and reduction of hydrogen peroxide on metal-supported magnesia is thermodynamically favorable.

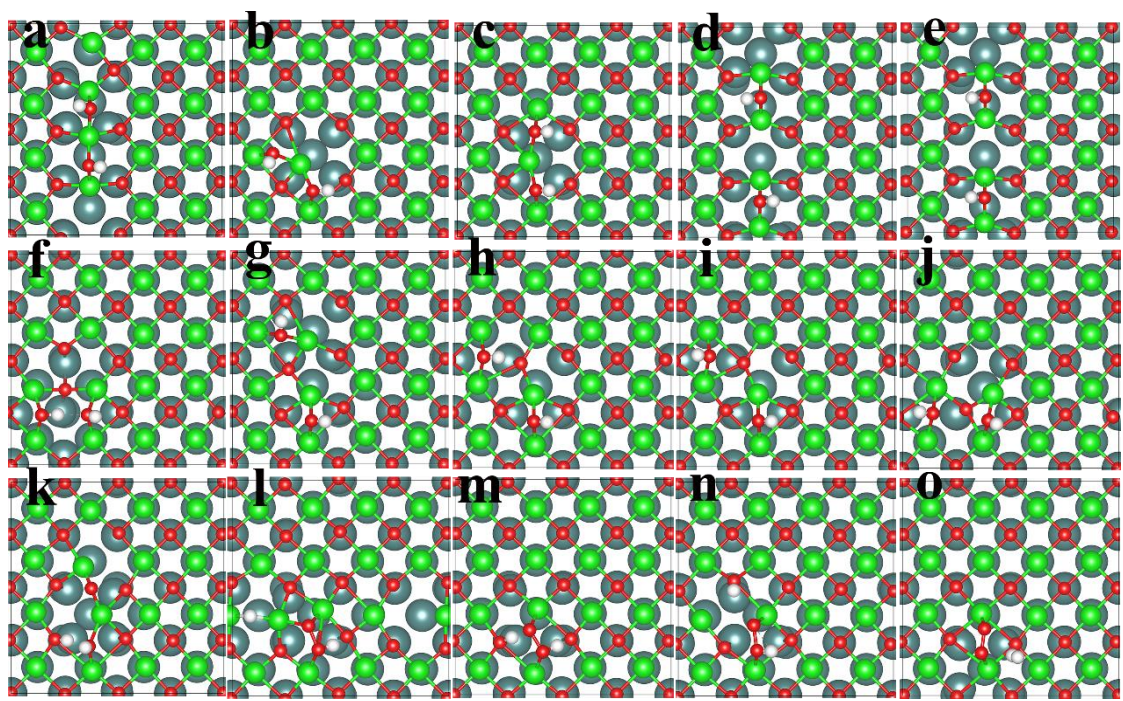

**Figure 11.** Various dissociative adsorption structures of hydrogen peroxide on 1 ML MgO(001)/Mo(001) surface.

**Table 4.** The relative energies, adsorption energies (in eV) and structural features for dissociative adsorption of hydrogen peroxide adsorbing on 1 ML – 10 ML MgO(001)/Mo(001)[a]

| Structure | Relative energy | Adsorption energy | Structural feature |
|---|---|---|---|
| 11a | 0 | -6.82 | 2 hydroxyls |
| 11b | 0.42 | -6.40 | 2 hydroxyls |
| 11c | 0.42 | -6.40 | 2 hydroxyls |
| 11d | 0.58 | -6.24 | 2 hydroxyls |
| 11e | 0.59 | -6.24 | 2 hydroxyls |
| 11f | 0.64 | -6.18 | 2 hydroxyls |
| 11g | 0.75 | -6.07 | 2 hydroxyls |
| 11h | 0.93 | -5.89 | 2 hydroxyls |
| 11i | 1.01 | -5.81 | 2 hydroxyls |
| 11j | 1.54 | -5.29 | 2 hydroxyls |
| 11k | 2.10 | -4.72 | water and oxygen |
| 11l | 4.63 | -2.19 | hydride and OOH |
| 11m | 4.82 | -2.00 | $O_sH$ and OOH |
| 11n | 4.82 | -1.10 | $O_sH$ and OOH |
| 11o | 5.72 | -0.85 | superoxygen and $H_2$ |

[a] the corresponding structures are shown in Figure 11.

## 3.7 Coadsorption of hydrogen peroxide and solvent (water/methanol) molecules on ultrathin magnesia film

The interaction of hydrogen peroxide with ultrathin magnesia film covered with water and methanol is considered to investigate the influence of solvent molecules on the dissociation and reduction behavior of hydrogen peroxide, as shown in Figure 11. When the magnesia film is covered with a monolayer water molecules, four water molecules dissociates to form surface hydroxyls (as shown in Figure S2). As in the first case (Figures 12a and 12b), the hydrogen peroxide approach the dissociated water at the solvated surface. $O_w1H$, the dissociation product of water, is further split by hydrogen peroxide, which produces a new water molecule ($H_2O1$). The $O_w1$ grasps hydrogen from adjacent water ($H_2O_w3$) to form new surface hydroxyls $O_w1H$ and $O_w3H$. Another fragment of hydrogen peroxide O2H reacts with the molecular water $H_2O_w2$ to form a new water $H_2O2$, and surface hydroxyl group $O_w2H$. The dissociation reaction release large amount of energy (-6.89 eV) and adds two more hydroxyl groups on the magnesia films, although the hydrogen peroxide doesn't contact with the magnesia directly (because of the solvent layer). The dissociation products, two newly formed water molecules doesn't adsorb directly on the surface magnesium, but they are stabilized by the strong hydrogen bonds surrounding them, with bond distances of 1.333 Å, 1.445 Å and 1.535 Å, for d(HO1H…$O_w$1), d($H_2O$2…H) and d(HO2H…$O_w$2H), respectively. In addition, the $H_2O$1 forms two hydrogen bonds with bond distances 1.671 Å and 1.752 Å. As in the second case (Figures 12c and 12d), hydrogen peroxide approaches the surface from the nondissociative water molecules. The dissociation product HO1 forms an anion, whose bader charge is calculated to be -0.686 e. This hydroxyl anion is stabilized by four HO1…$H_2O$ hydrogen bonds, with bond distances 1.592 Å, 1.594 Å, 1.623 Å and 1.670 Å, respectively. The HO2 reacts with surface hydroxyl $O_w$H to form a new water molecule $H_2O$2, and subsequently $O_w$ grasps hydrogen from adjacent water to form a new hydroxyl. The newly formed water $H_2O$2 is stabilized by three hydrogen bonds, with bond distances d($H_2O$2…HOH) = 1.655 Å, 1.704 Å, d(HO2H…$O_w$H) = 1.324 Å. For the second case, the reaction is spontaneous with large dissociative adsorption energy of -7.514 eV, and yields two hydroxyl groups, which are surrounded by hydrogen bonding interaction from solvent molecules. Moreover, we change the water to methanol to investigate the influence of solvent category on the reduction and dissociation behavior of hydrogen peroxide. Because of the larger volume of methanol than water, after optimization, two methanol molecules get squeezed out of the surface and do not bind with surface magnesium any more (Figures 13 and S3). The hydrogen peroxide is split spontaneously on magnesia film covered with a monolayer of methanol, with large dissociative adsorption energy of 4.686 eV. After the peroxide bond is destroyed, one hydroxyl of hydrogen peroxide rotates to reversed direction. The oxygen end of the rotated hydroxyl seizes the hydrogen of methanol to form a water molecule, and the hydrogen end of the rotated hydroxyl forms strong hydrogen bonding interaction (with OH…OH bond distance 1.252 Å) with another hydroxyl of dissociative hydrogen peroxide. In actual system, hydrogen peroxide and solvent molecules can be coadsorbed directly on the ultrathin magnesia surface, as shown in Figure 14. The dissociative hydrogen peroxide on ultrathin magnesia film covered with submonolayer water is isostructural with single molecular hydrogen peroxide dissociatively adsorbing on bare MgO(001)/Mo(001) surface. The reduction and dissociation reaction on surface covered with submonolayer water is calculated to be highly exothermic by -6.884 eV.

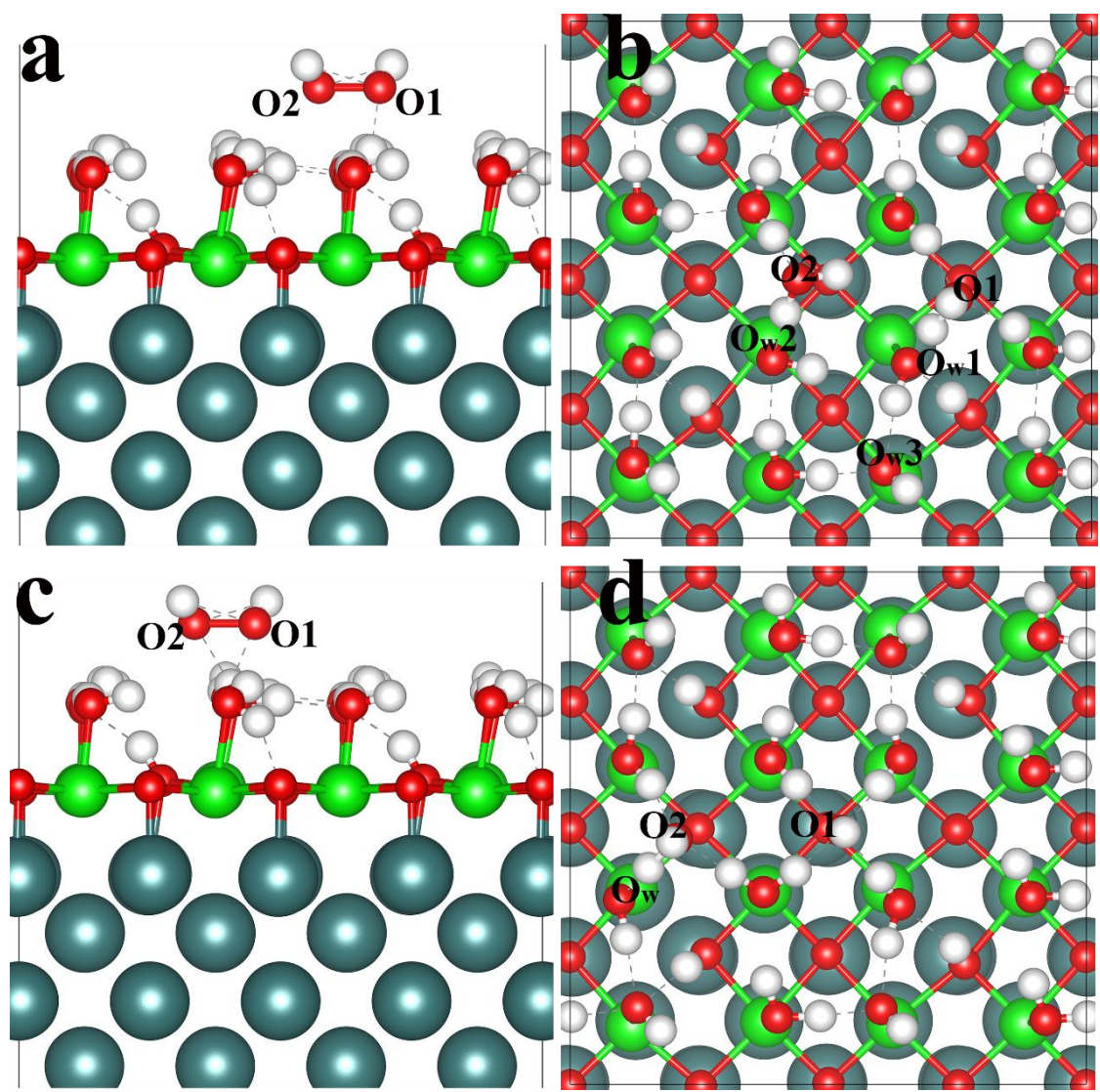

**Figure 12.** Coadsorption of hydrogen peroxide and water molecules (one monolayer) on 1 ML MgO(001)/Mo(001). (a) (c) Typical adsorption sites; (b) (d) optimized dissociative adsorption structures.

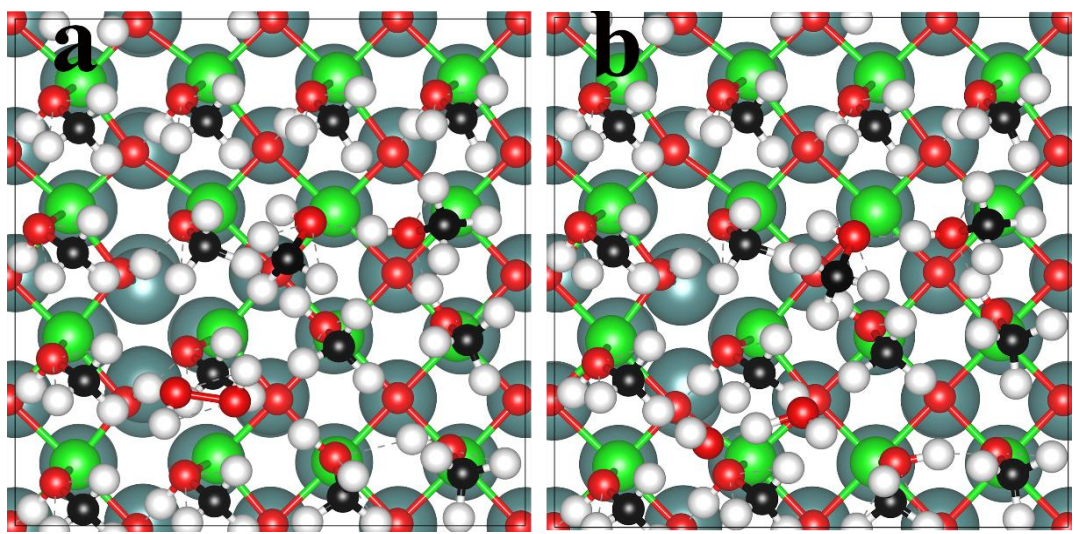

**Figure 13.** Coadsorption hydrogen peroxide and methanol molecules (one monolayer) on 1 ML MgO(001)/Mo(001). (a) Top view; (b) side view.

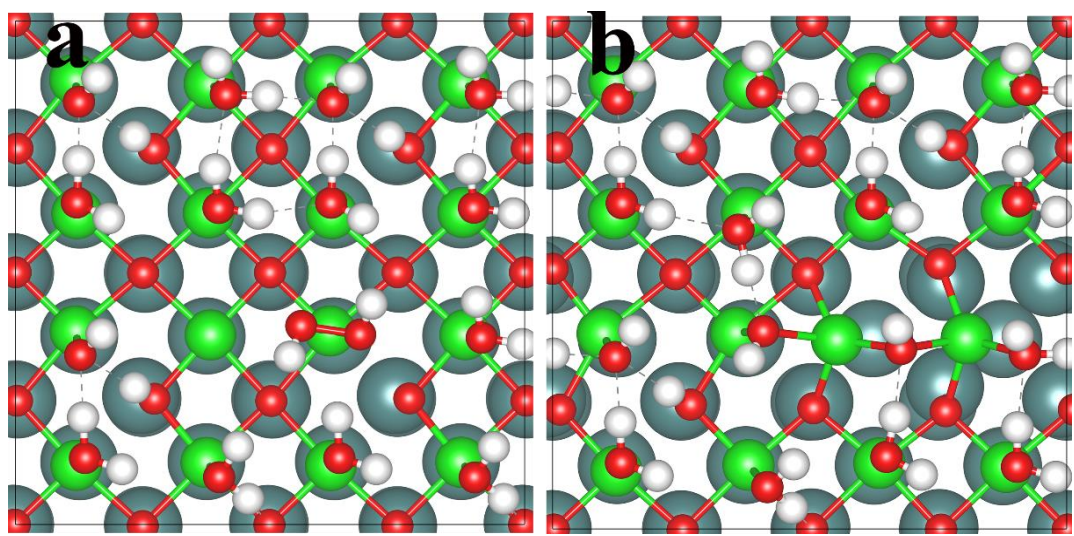

**Figure 14.** Coadsorption of hydrogen peroxide and water molecules (submonolayer) on 1 ML MgO(001)/Mo(001). (a) Initial structure; (b) optimized dissociative adsorption structure.

### 3.8 Fragmentation and reduction of peroxide bonds of diethyl peroxide and peroxyacetone

In order to confirm the catalytic effect of ultrathin magnesia toward reducing peroxide bond, we further perform density functional calculations to investigate the adsorption behavior of diethyl peroxide and peroxyacetone on bulk magnesia (001) and ultrathin magnesia (001) surface deposited on metal substrate. The optimized molecular adsorption and dissociative adsorption structures are illustrated in Figures

15a – 15g, and the energetic properties are shown in Table 5. Diethyl peroxide can adsorb molecularly on bare magnesia (001), releasing heat of -0.563 eV (Table 5). However, the dissociation reaction of diethyl peroxide on bare magnesia (001) is highly endothermic, with positive dissociative adsorption energy of 0.812 eV. If the reaction initiates from the molecular adsorption state, the fragmentation process of diethyl peroxide on bare magnesia require absorbing energy larger than 1.375 eV. The dissociation product is ethoxyl group ($C_2H_5O$-). Peroxyacetone adsorbs on bare magnesia (001) with energy release of -0.189 eV, much less than that of diethyl peroxide. The quite weak adsorption of peroxyacetone on bare magnesia (001) can be attributed to the steric hindrance of the substituting isopropylidene group -$(CH_3)_2C$-, which obstructs the direct binding interaction between peroxide group with surface magnesium. Peroxyacetone can dissociate to three $(CH_3)_2C(O-)_2$ groups, with significantly smaller energy elevation of 0.074 eV, as compared with that of diethyl peroxide, indicating the relative instability of peroxyacetone. In reality, the peroxyacetone is a non-nitrogenous explosive, which is more difficult to detect than nitrogenous explosives, and can explode if subjected to heat, friction, or shock. The slight energy difference between the molecular adsorption state and dissociative adsorption state is not enough for appropriate handling of this explosive to avoid the accidental detonation. The molecular adsorption state of peroxyacetone present negative adsorption energy -0.428 eV. However, the molecular adsorption state of diethyl peroxide does not exist, indicating the high chemical activity of ultrathin magnesia. When peroxyacetone is adsorbed on metal-supported magnesia (001), the peroxyacetone dissociates to $(CH_3)_2C(O-)_2$ species smoothly, and the process is exothermic by -14.055 eV relative to its isolate state. The metal supported magnesia (001) can split diethyl peroxide spontaneously with energy release of -5.174 eV. The dissociative adsorption energy of peroxyacetone is much larger than that of diethyl peroxide, because all three peroxide bonds participate in the exothermic fragmentation process.

Essentially, the diethyl peroxide and peroxyacetone can be seen as derivatives of hydrogen peroxide, with substituent group (SG) ethyl and isopropylidene. At dissociative adsorption state, the ethyl charges on bare magnesia (001) and metal-supported magnesia (001) are calculated to be +0.561 e and +0.451 e (as listed in Table 6), respectively. While the $O_{dis}$ charges are calculated to be -1.026 e and -1.300 e on bare and metal-supported magnesia, respectively, indicating the ethyoxyl grasps more electrons from the metal-supported magnesia. For peroxyacetone dissociation on bulk magnesia (001), the isopropylidene carries charge of +1.525 e, averagely, which is much larger than that of ethyl group, for the isoproylidene is connected chemically to two oxygen atoms. The $O_{dis}$ carries average charge of -1.052 e and the dissociation of peroxyacetone on bulk magnesia leads to the positively charged magnesia surface (+1.734 e). The three oxidizing species $(CH_3)_2C(O-)_2$ form after peroxyacetone dissociation on metal-supported magnesia (001), and the electrons reserved in molybdenum substrate are substantially less than the dissociation of diethyl peroxide. In addition, the peroxyacetone dissociation on metal-supported magnesia results in the severe oxidation of magnesia film, with positive charge of +9.670 e. The substituent group isoproylidene carries average charge +0.959 e, less than that of the dissociative state at bulk magnesi. Moreover, after dissociation of peroxyacetone on ultrathin magnesia, the $O_{dis}$ carries charge of 1.307, larger than that on bulk magnesia, indicating the more effective charge transfer and chemical binding interaction between $O_{dis}$ and ultrathin magnesia.

The differential charge density contours are depicted in Figures 15h-15n. We could observe vividly the substituent group ethyl abstracts electrons from the oxide-metal hybrid structure (Figure 15h). Compared with the isolated ethyoxyl, the electron density in C-C bond of this group increases slightly. This result can be ascribed to the fact that the $O_{dis}$ mainly obtain electrons from ethyl group, while after adsorbed on metal-supported magnesia, $O_{dis}$ could grasp considerable electrons from both ethyl and the surface. When diethyl peroxide is adsorbed molecularly on bulk magnesia (001), the magnesia transfers a small quantity of electrons to peroxide group. Simultaneously, due to the inducing effect of surface oxygen, the weak $O_s…H$ hydrogen bonding interaction leads to the further electron shift from hydrogen to carbon (Figure 15i). The dissociation of diethyl peroxide on bulk magnesia (001) destroy the peroxide bond, and the two $O_{dis}$ are connected with two surface magnesium, respectively. It can be observed that the ethyoxyl accumulates electrons from the hybrid surface, and the surface reaction sites experience electron depletion (Figure 15j). Analogous to the dissociative adsorption on metal-supported magnesia, the electron density between C-C increases compared with the isolated ethyoxyl group, after dissociative adsorption of diethyl peroxide on bulk magnesia (001). For the dissociation reaction on bulk magnesia (001), the differential charge density contour mainly distribute in small areas at the reaction sites. Whereas, the large amount of charge transfer between ultrathin magnesia and the oxidizing adsorbates lead to the large distributing areas of differential charge density. This result is also related to the structural characteristics. The dissociation of diethyl peroxide on bulk magnesia (001) does not results in the destruction of Mg-O ionic bonds, and the obvious severe structure relaxation occurs on metal-supported magnesia, which destroys several Mg-O ionic bonds at the reaction sites. For the adsorption of peroxyacetone, the substituting group isoproylidene is much larger than ethyl, which screens the binding interaction between O-O species and the surface magnesium, and causes less charge transfer of metal-supported or bulk magnesia to adsorbing molecule, as compared with molecular adsorption of diethyl peroxide on bare magnesia (as shown in Figures 15k and 15m). Compared with the isolated molecule, the adsorbed peroxyacetone can form C-H…$O_s$ hydrogen bond, and the shared electrons between C-H shift further to carbon. As shown vividly in Figure 15l and 15n, when peroxyacetone is fragmented on metal-supported and bulk magnesia (001), three $(CH_3)_2C(O-)_2$ accumulate electrons from surface indicating the strong chemical interaction between adsorbates and surface. The $(CH_3)_2C(O-)_2$ grasps electrons more broadly from metal-supported magnesia than bulk magnesia, indicating the strong ability of metal-supported magnesia for splitting and reducing peroxyacetone.

**Table 5.** The adsorption energies (in eV) of diethyl peroxide and peroxyacetone on bulk MgO(001) and 1 ML MgO/Mo(001).

| Surface | State | Diethyl peroxide | Peroxyacetone |
|---|---|---|---|
| Bulk MgO(001) | molecular | -0.563 | -0.189 |
|  | dissociative | 0.812 | 0.074 |
| MgO/Mo(001) | molecular | — | -0.428 |
|  | dissociative | -5.174 | -14.055 |

**Table 6.** Charge distributions of oxygen in dissociative peroxide group ($O_{dis}$), substituting group on hydrogen peroxide (SG), MgO film and Mo substrate based on Bader charge analysis (unit in electron), for diethyl peroxide (DP) and peroxyacetone (PA) dissociative adsorption on bulk and ultrathin MgO (001).

| Surface | Adsorbate | $O_{dis}$ | SG | MgO | Mo |
|---|---|---|---|---|---|
| bulk MgO (001) | dissociative DP | -1.026 | +0.561 | +0.930 | — |
|  | dissociative PA | -1.052 | +1.525 | +1.734 | — |
| MgO /Mo(001) | dissociative DP | -1.300 | +0.451 | +8.836 | -7.139 |
|  | dissociative PA | -1.307 | +0.959 | +9.670 | -4.704 |

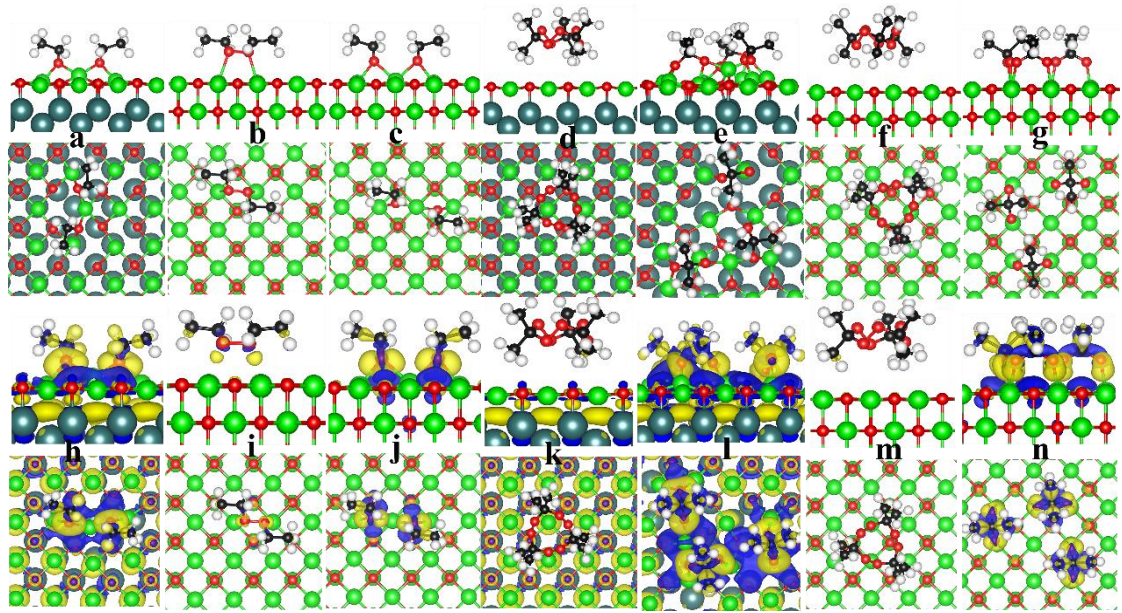

**Figure 15.** The side and top structures (a), and differential charge density (h) for dissociative adsorption of diethyl peroxide on 1 ML MgO(001)/Mo(001). The side and top structures (b), and differential charge density (i) for molecular adsorption of diethyl peroxide on bulk MgO(001). The side and top structures (c), and differential charge density (j) for dissociative adsorption of diethyl peroxide on bulk MgO(001). The side and top structures (d), and differential charge density (k) for molecular adsorption of peroxyacetone on 1 ML MgO(001)/Mo(001). The side and top structures (e), and differential charge density (l) for dissociative adsorption of peroxyacetone on 1 ML MgO(001)/Mo(001). The side and top structures (f), and differential charge density (m) for molecular adsorption of peroxyacetone on bulk MgO(001). The side and top structures (g), differential charge density (n) for dissociative adsorption of peroxyacetone on bulk MgO(001). The differential charge density is obtained by subtracting charge density of the adsorbed molecules (or dissociated products), the MgO film and the molybdenum substrate from the whole system. The isosurface value is set to be 0.003 e bohr$^{-3}$. The blue and yellow colors stand for the electron loss and electron gain, respectively.

## 4. Conclusions

In summary, the hydrogen peroxide dissociation on MgO(001) films deposited on transition metal surface is uncovered for the first time by performing periodic density-functional theory methods with long-range dispersion correction. The pristine MgO(001) surface showing chemical inertness is extremely difficult to react with hydrogen peroxide. The hydrogen peroxide is dissociated smoothly and reduced to surface hydroxyls on perfect stoichiometric magnesia (001) films deposited on transition metal substrates. Drastically different from the dissociation structure on bare magnesia (001) with positive adsorption energy 0.774 eV, the dissociative adsorption energies of all the considered fragmentation configurations on 1 ML MgO (001)/Mo are substantially negative, indicating that dissociation and reduction of hydrogen peroxide on metal-supported magnesia is thermodynamically favorable. It can be deduced that substrate effect should play an important role in dissociation and reduction of hydrogen peroxide according to the comparison amongst the dissociation behavior on bare magnesia, extended bare magnesia, and metal-supported magnesia. The dissociative adsorption energy decreases monotonously with increasing film thickness, which indicates the lower reactive activity of thick oxide films. In addition, we examined the suitability of several transition metal slabs (molybdenum, silver, vanadium, tungsten and gold) combined with insulating oxide for splitting hydrogen peroxide. The results indicate that the dissociative adsorption energies become much larger with the increase of lattice constants of substrate slabs. Thus, the chemisorption strength could be tuned by the category of metal slabs as well as the thickness of oxide film. The reactivity enhancement for energetically and dynamically favorable decomposition of hydrogen peroxide on metal-supported oxide films is interpreted by characterizing the geometric structures, Bader charges, density of states, electron localization function, differential charge densities and particular occupied orbitals. The interaction of hydrogen peroxide with ultrathin magnesia film covered with water or methanol is investigated to reveal the influence of solvent molecules on the dissociation and reduction behavior of hydrogen peroxide. To uncover the catalytic activity of ultrathin magnesia toward splitting organic peroxides, the fragmentation and reduction of diethyl peroxide and peroxyacetone are also studied on oxide-metal composite nanostructure. We anticipate that the results here can provide useful clue for detecting and reducing hydrogen peroxide and organic peroxides by employing nanostructured oxide insulators.

ASSOCIATED CONTENT

**Appendix A. Supplementary data**

Supplementary data associated with this article can be found, in the online version. Maximum and minimum bond distances in $z$ direction for hydrogen peroxide on bulk and ultrathin magnesia (001) (Figure S1); Adsorption structure of a monolayer water molecules (Figure S2); Adsorption structure of a monolayer methanol molecules (Figure S3); Adsorption energies and total energies with increasing oxide thickness, for dissociative hydrogen peroxide adsorbing on 1 ML – 10 ML MgO(001)/Mo(001) (Table S1).

AUTHOR INFORMATION

**Corresponding Authors**

*(B.Z.) E-mail: zhaobin@nankai.edu.cn

*(P.C.) E-mail: pcheng@nankai.edu.cn

**Acknowledgments**

This work was supported by the NSFC (Grants 21625103, 21571107, and 21421001), Project 111 (Grant B12015), and the SFC of Tianjin (Grant 15JCZDJC37700). The density functional calculations in this research were performed on TianHe-1(A) at National Supercomputer Center in Tianjin.

**Graphical abstract**

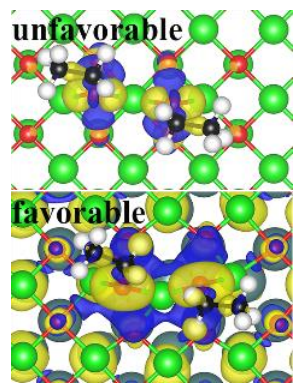